\documentclass[sigconf]{acmart}

\AtBeginDocument{%
  }

\setcopyright{rightsretained}
\copyrightyear{2024}
\acmYear{2024}
\acmDOI{XXXXXXX.XXXXXXX}

\acmConference[arXiv]{arXiV}{July 9, 2024}{}

\acmISBN{978-1-4503-XXXX-X/18/06}

\usepackage{pythonhighlight}

\usepackage{array}
\begin{document}
\title{SEBA: Strong Evaluation of Biometric Anonymizations}

\author{Julian Todt}
\affiliation{%
 \department{KASTEL Security Research Labs}
 \institution{Karlsruhe Institute of Technology}
 \streetaddress{Am Fasanengarten 5}
 \city{Karlsruhe}
 \state{}
 \country{Germany}
 \postcode{76131}
}
\email{julian.todt@kit.edu}

\author{Simon Hanisch}
\affiliation{%
 \department{Centre for Tactile Internet (CeTI)}
 \institution{Technical University Dresden}
 \streetaddress{Nöthnitzer Str. 46}
 \city{Dresden}
 \state{}
 \country{Germany}
 \postcode{01187}
}
\email{simon.hanisch@tu-dresden.de}

\author{Thorsten Strufe}
\affiliation{%
 \department{KASTEL Security Research Labs}
 \institution{Karlsruhe Institute of Technology}
 \streetaddress{Am Fasanengarten 5}
 \city{Karlsruhe}
 \state{}
 \country{Germany}
 \postcode{76131}
}
\email{thorsten.strufe@kit.edu}
\begin{abstract}
Biometric data is pervasively captured and analyzed.
Using modern machine learning approaches, identity and attribute inferences attacks have proven high accuracy.
Anonymizations aim to mitigate such disclosures by modifying data in a way that prevents identification.
However, the effectiveness of some anonymizations is unclear.
Therefore, improvements of the corresponding evaluation methodology have been proposed recently.
In this paper, we introduce SEBA, a framework for strong evaluation of biometric anonymizations.
It combines and implements the state-of-the-art methodology in an easy-to-use and easy-to-expand software framework. This allows anonymization designers to easily test their techniques using a strong evaluation methodology.
As part of this discourse, we introduce and discuss new metrics that allow for a more straightforward evaluation of the privacy-utility trade-off that is inherent to anonymization attempts.
Finally, we report on a prototypical experiment to demonstrate SEBA's applicability.
\end{abstract}



\maketitle

\section{Introduction}

Social networks, media, smart cities, the \textit{metaverse} ---  all these applications imply the collection of vast amounts of personal data, much of it of biometric nature. Biometric traits like face, voice, or gait are widely shared, creating privacy issues for recorded individuals.
Privacy risks arise when biometric data is used to infer personal information, particularly the identity of the recorded individual.
With recent advances in deep learning, new recognition methods have emerged that are capable of inferring the identity of individuals with unprecedented accuracy.
One promising option to mitigate these privacy risks is anonymization.
This means that the recorded data is modified by an anonymization method in such a way that identification shall no longer be possible while preserving some the utility of the data for its original processing purpose.

An astounding number of anonymization methods have been proposed lately, particularly to prevent face recognition in images, and have been evaluated with impressive claims.
However, most recently, problems with the common evaluation methodology for anonymizations have been highlighted.
They show a weak attacker model and a lack of reversibility evaluation lead to a false sense of privacy \cite{hanisch_false_2024,todt_fantomas_2024}.
Consequently, improvements to the methodology have been proposed by Hanisch et al.~\cite{hanisch_false_2024} and Todt et al.~\cite{todt_fantomas_2024}.
These include the reduction of the identity set and the estimation of reversibility (de-anonymization) of the anonymization.
What is still missing is both a combination of these two methodologies and a comprehensive software framework that implements it.
Further, it should be easily extendable to allow for more anonymizations of more biometric traits to be evaluated using this methodology.
Finally, a fair comparison of anonymizations also requires an evaluation of the privacy-utility trade-off which is also still lacking.

In this paper, we present \emph{SEBA}, a software framework for \emph{Strong Evaluation of Biometric Anonymizations}.
SEBA implements the state-of-the-art evaluation methodologies for biometric anonymizations in an easy-to-use and easy-to-extend python framework.
SEBA automatically manages and caches -- when appropriate -- datasets, performs anonymization, de-anonymization, and recognition and calculates and visualizes metrics.
We include implementations for a variety of common (de-)anonymizations and recognition methods for both face images and gait sequences.
Extendability was a priority during design, so while SEBA is ready-to-go with the provided methods, own (de-)anonymizations, recognition methods, traits and datasets can be easily added.
SEBA is available at \url{https://github.com/kit-ps/seba}.

Our contribution in this paper is three-fold:
\begin{itemize}
    \item A simplified and combined evaluation methodology based on the state-of-the-art
    \item SEBA, the first, easy-to-use and easy-to-extend software framework to evaluate biometric anonymizations
    \item Metrics and visualizations that allow for a straightforward evaluation and comparison of the privacy-utility trade-off of anonymizations.
\end{itemize}

\section{Background}\label{sec:background}
In this section we provide an overview over the underlying concepts of SEBA.

\subsection{Biometrics, Identification \& Anonymization}
Biometric traits are physiological or behavioral characteristics of humans, examples of them are fingerprints, face, voice and gait.
With the exception of so-called soft biometrics, they are permanent and distinctive and can therefore be used to identify individuals.
Biometric recognition systems are able to infer personal information from instances of these traits.
These inferences can be attributes (e.g. sex, age, weight) or the identity of the individual.
Because biometric data is collected ubiquitously in social networks, media, smart cities, the metaverse, and more, the possibility of these inferences raises privacy concerns.

One possible mitigation of biometric recognition is anonymization which has two goals:
On the one hand, to irreversibly remove all identifying information from the data, so that identification is no longer possible.
And on the other hand, to preserve the utility of the data, so that it can still be used for its intended use-case.
Oftentimes, these two goals are contrary to each other, and a trade-off needs to be made with many anonymizations having parameters to adjust it.

\subsection{Recent Evaluation Methodology Advances}
Hanisch et al. \cite{hanisch_false_2024} show that many evaluations of anonymizations use a weak attacker model, inspired by the evaluation of recognition methods.
However, a good evaluation of anonymizations should assume a strong attacker, since anonymizations ought to protect everybody, incl. potentially more vulnerable groups, and not just the majority of users.
Therefore, they propose to (pre-)train recognitions on anonymized data, test different recognitions, and reduce the number of identities in the evaluation using specific strategies in order to create a diverse data set which is a challenge to anonymize.

Todt et al. \cite{todt_fantomas_2024} investigate the reversibility of face anonymizations and show that a majority of methods are at least partially reversible.
They therefore propose to use a general de-anonymization using machine learning to reverse the anonymization before attempting to re-identify individuals as a step when evaluating anonymizations.

\section{Related Work}\label{sec:relatedwork}
In 2020, the first VoicePrivacy Challenge was held to compare different speaker anonymizations using a common methodology \cite{tomashenko_introducing_2020}.
The challenge provided an evaluation framework to help the speaker anonymization community compare their anonymizations.
Unlike the VoicePrivacy Challenge's framework, our framework can be used for biometric anonymization for any trait and implements additional methods for testing anonymizations, such as identity set reduction and reconstruction of the original data.

\section{Motivation and Goals}\label{sec:goals}
While many anonymizations have been proposed in the past, they are difficult to compare since their evaluation strategies vary significantly.
At the same time, the used evaluation methodologies are also often lacking, for example by assuming a weak attacker that is unaware of the anonymization.
The most recent advances to the evaluation methodology make for a more rigourous evaluation, but without an easy-to-use software framework that combines and implements them, their use might be limited.
Therefore, our goals with SEBA are as follow:
\begin{itemize}
    \item Define common evaluation scenarios and attackers for biometric anonymization.
    \item Introduce a rigorous evaluation of biometric anonymization techniques.
    \item Establish a fair comparison of the privacy-utility trade-off of biometric anonymizations.
    \item Provide a flexible open-source software framework which can be easily extended by other anonymization designers to test their techniques and to incorporate new evaluation methodologies.
\end{itemize}

\subsection{Adversary Model}\label{sec:attacker}
We assume scenarios where a person shares their biometric data to gain utility. Examples include posting face images on social media, using a wearable to monitor health data and sending it to an online service for analysis, speaking in a voice chat, or using motion capture to animate an avatar in the metaverse. To protect themselves from unwanted privacy inferences from their data, they use an anonymization to protect their personal information.

We assume an attacker who wants to identify a person from the shared biometric data. The attacker has access to the shared anonymized biometric data as well as an additional biometric data set on which to train a biometric system. Furthermore, the attacker knows that the data has been anonymized and which anonymization was used with which parameters. The attacker does not have access to any secrets used in the anonymization process.

\section{Evaluation Methodology}
In the following we present our evaluation methodology on which our evaluation framework is based.
For this, we combine the methodologies introduced by Hanisch et al. and Todt et al. \cite{hanisch_false_2024, todt_fantomas_2024}.

\subsection{Rigorous Privacy Evaluation}
As mentioned in the background, an anonymization must protect all of its users, especially potential outlier as they may be more vulnerable.
Therefore, a worst-case scenario should be used with a strong attacker model.
The methodology must evaluate whether anonymization is capable of protecting the identities of individuals even in challenging scenarios.

Our attacker performs identification on a closed-set, i.e. the possible suspects are known and the attacker selects the identity with the highest probability. Further, the attacker has multiple recognition models available, and can select the one which performs the best on the given anonymization.

Due to the availability of an additional biometric data set and the knowledge of the anonymization, the attacker is able to train the biometric recognition model with anonymized data. As multiple previous works \cite{newton_preserving_2005, hanisch_false_2024} have shown this is an effective way to increase identifiability because the recognition model can adapt to the anonymization.
Therefore, a strong attacker will train and enroll their system with anonymized data.

As introduced by Hanisch et al. \cite{hanisch_false_2024}, in our methodology the number of identities is reduced to create a more challenging data set to anonymize.
We employ the author's distinctive selection strategy which selects identities which are easy to distinguish. %
This strategy uses a recognition model trained on anonymized data to encode each data point into an identity space.
There, identities are selected which have all their data points close together and far away from data points of other identities.

We also add the de-anonymization proposed by Todt el al. \cite{todt_fantomas_2024}.
There, the attacker trains a general de-anonymization machine learning model using the additional data set that they have available and a corresponding anonymized data set.
Then, before the selected data set is attempted to be re-identified, the images are de-anonymized using this model.
As there is no obvious indication which of the recently introduced methodologies might be better, we include de-anonymization as an optional step.
In the case of de-anonymization, we also train and enroll the recognition using clear (not anonymized) data as recommended by the authors.

\subsection{Measuring Utility}
Besides measuring the privacy via recognition performance, we also evaluate the utility which the anonymized data retains.
The utility depends on the use-case and as such a wide variety of utility measures exist, they include for example attribute and landmark similarity as well as naturalness.
While naturalness of face images is often evaluated using user studies, more recently face detections have been used \cite{hukkelas_deepprivacy_2019} .
The rationale is that if a detection method that was trained on natural images is able to detect a face with high confidence, it must appear fairly natural.

\section{Fair Comparison \& Metrics}
There is no single metric for the privacy-utility trade-off, which makes comparing anonymizations difficult.
Rather, both privacy and utility have their own metrics.
For example privacy is usually measured using the (balanced) accuracy of the recognition.
Using multiple metrics, however, is problematic since most anonymizations can be parameterized in such a way to achieve good results in either, but not both at the same time.
Therefore, to fairly compare different anonymizations, their privacy-utility trade-off must be measured using a metric that combines privacy and utility.
We propose to measure both privacy and utility at multiple different parameterizations that ideally span a wide range.
These can then be plotted with privacy on the x-axis and utility on the y-axis.
This not only allows for an easy visual comparison of anonymizations, but by computing the area under curve (AUC), we can generate a metric that combines privacy and utility.
This allows for a fair comparison of anonymizations.
A few anonymizations do not allow for a straightforward parameterization of their privacy-utility trade-off.
In these cases, we simply consider their single data point and otherwise a rectangle defined by the data point and the origin of the coordinate system.

Finally, whenever considering metrics for anonymizations, it should be noted in which range they are in.
For privacy, chance level, i.e. randomly guessing identities, is the best an anonymization can achieve, while clear level, i.e. without anonymization, is the worst.
For utility, the range will depend on the specific utility measure used.

We recommend the AUC to be computed both with and without de-anonymization for a relevant number of parameter choices.
As in the rest of the evaluation, a worst-case scenario should be assumed.
Therefore, the lowest value across the two variants for each anonymization should be used to compare them.

\section{Design}\label{sec:design}
\begin{figure}[h!]
    \centering
    \includegraphics[width=0.45\textwidth]{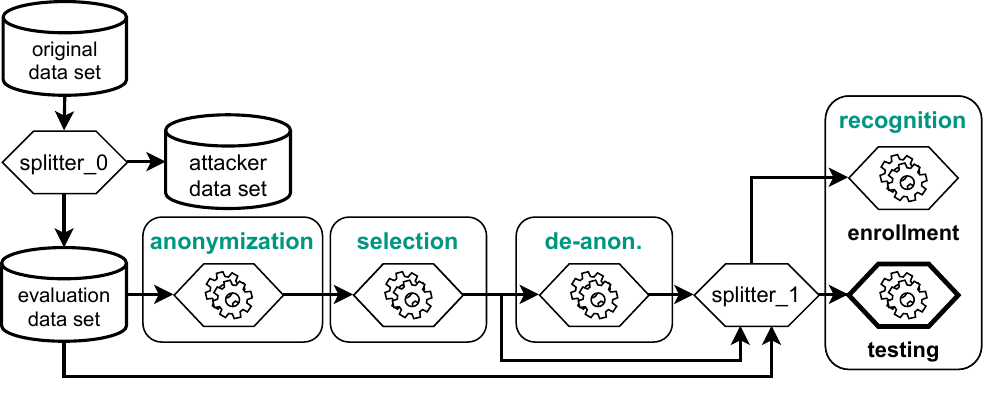}
    \caption{Design of our evaluation framework (simplified).}
    \label{fig:systemdesign}
\end{figure}

In this section, we present the design of our software framework, SEBA, that implements the evaluation methodology described above.
To facilitate its use in future anonymization evaluations, a focus in its design was extendability for future use-cases.
This is why biometric traits, anonymizations, selection strategies, de-anonymizations and recognition methods are easily replaceable, and new ones simply need to implement a well-defined interface to work with SEBA. An overview of the interfaces can be found in Appendix~\ref{sec:interfaces}.
At the same time, SEBA is directly usable with a significant number of techniques already implemented. See Appendix~\ref{sec:implemented} for a list.
A simplified overview over the design can be found in Figure~\ref{fig:systemdesign}.

When starting the evaluation of an anonymization, the original data set is first split into two parts with disjoint identities: the attacker data set, which contains the data used by the attacker to train its recognition and de-anonymization model, and the evaluation data set which is used to evaluate the anonymization.
The evaluation data is first anonymized and then the selection strategy reduces the number of identities in the set.
Then, as an optional step, the selected set is de-anonymized.
Finally, the evaluation data set is split for the recognition system into one part which is used for the enrollment and one part for the testing.

Before each step, SEBA checks whether the exact data set already exists from previous evaluation runs and then re-uses these sets when appropriate to save resources.

\section{Implementation}\label{sec:impl}
SEBA is implemented in Python (version 3.10) using the numpy (version 1.24.3) and scikit-learn (version 1.3.0) libraries.
Further libraries are conveniently packaged as a conda \cite{anaconda} environment.
Whenever possible, the author's original implementation was used.
To keep the initial footprint of SEBA small and due to license concerns, these implementations, as well as any data sets, are not included.
However, installation scripts which include potentially necessary patches or adapt the structure of data sets, are provided in the framework.
In general, a significant amount of related functionality is provided as scripts, for example generating configurations and plotting results.

\section{Sample Experiment}\label{sec:results}
In this section, we conduct an example experiment using SEBA and present its results.
We evaluate the commonly used anonymizations Gaussian blur and eye masking as well as the state-of-the-art method DeepPrivacy \cite{hukkelas_deepprivacy_2019}.
For each we run multiple parameter choices\footnote{Gaussian Blur: kernel (51, 99, 147, 195); Eye Masking: number of pixels from eye level to remove (50, 70, 90, 110, 130)} except for DeepPrivacy which does not allow easy parameterization.
Anonymized sample images can be found in Appendix~\ref{sec:samples}.
We use the common CelebA data set \cite{liu_deep_2015} which we pre-process to only include the face regions of images using the scripts in SEBA.
For selection, we use the distinctive strategy from Hanisch et al. \cite{hanisch_false_2024} to select 100 identities from the evaluation dataset.
As a utility metric, we use a face detection on the anonymized images and calculate the mean of the detector's confidence over all tested images.
If no face is detected, a confidence of zero is used.

\begin{figure}[h!]
    \centering
    \includegraphics[width=0.45\textwidth]{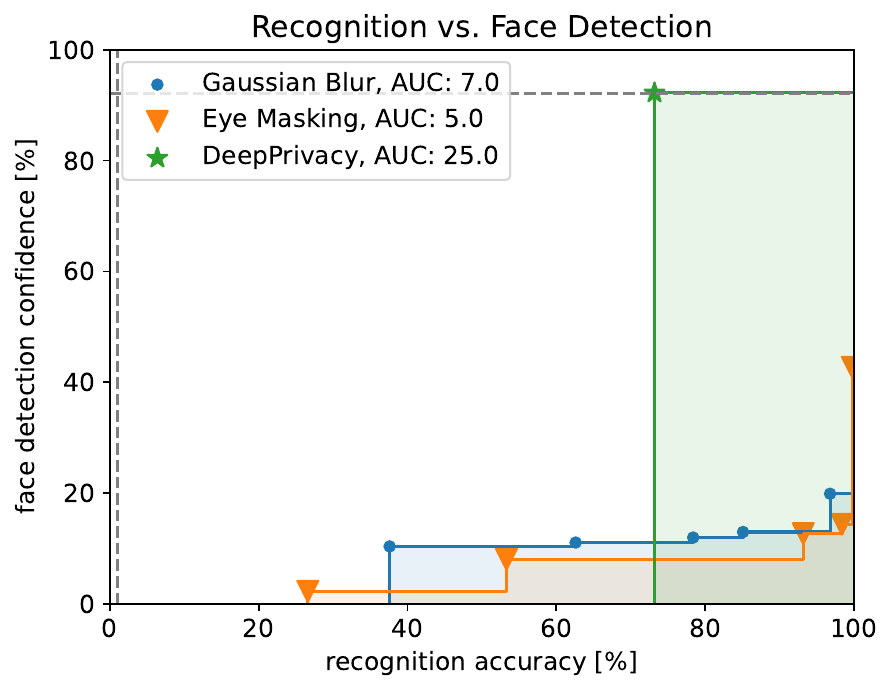}
    \caption{Results for privacy and utility for DeepPrivacy, Blurring and Eye Masking on 100 identities of CelebA \emph{without} de-anonymization. Clear-level utility and change-level privacy are indicated with dotted lines.}
    \label{fig:res-anon}
\end{figure}

\begin{figure}[h!]
    \centering
    \includegraphics[width=0.45\textwidth]{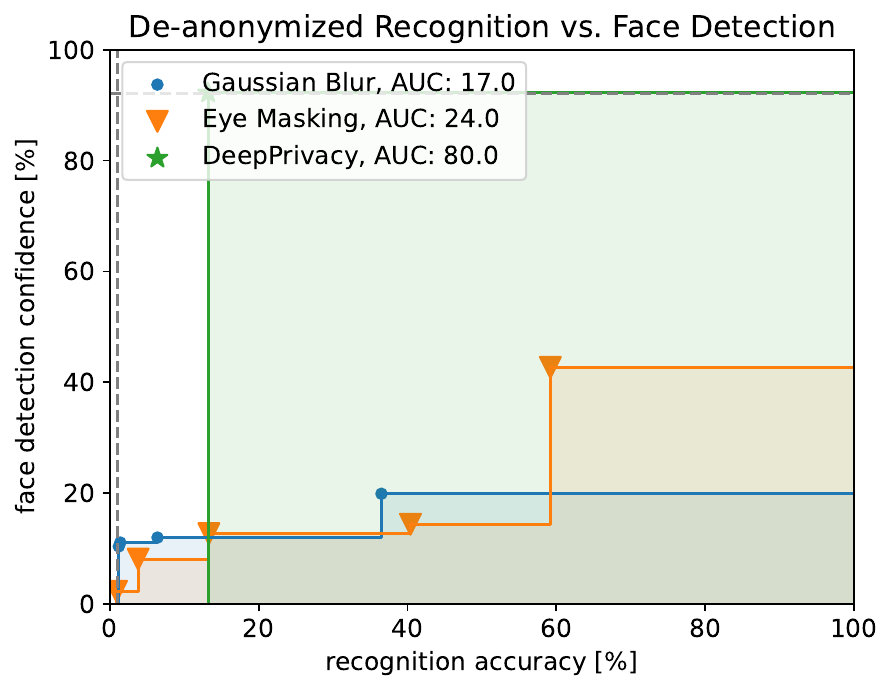}
    \caption{Results for privacy and utility for DeepPrivacy, Blurring and Eye Masking on 100 identities of CelebA \emph{with} de-anonymization. Clear-level utility and change-level privacy are indicated with dotted lines.}
    \label{fig:res-deanon}
\end{figure}

The results can be found in Figure~\ref{fig:res-anon} and Figure~\ref{fig:res-deanon} for the experiment without de-anonymization and with, respectively.
We find that DeepPrivacy achieves the highest face detection confidence, which is on par with the one on clear data (cmp. to horizontal dotted line).
This is because DeepPrivacy effectively replaces the anonymized face with a new synthetic one.
All other anonymizations achieve significantly lower utility.

For privacy, we find that high parameters for Gaussian blur and eye masking result in lower face recognition accuracies than DeepPrivacy, reducing the recognition rate to below 40\% or even 30\%.
At the same time, with lower parameters, while utility increases as expected, so does the recognition accuracies, up to close to clear-level.
For the experiment without de-anonymization, all measured accuracies are far away from the theoretical chance level that would be expected for anonymizations.

Comparing the two variants, we see that for the chosen anonymizations, the recognition achieves higher accuracies without de-anonymization and therefore the anonymizations have been evaluated more rigorously.
This is also shown by the AUC values which are lower for recognition on anonymized images.
At the same time, we already see differences between the two variants with Gaussian blur outperforming Eye Masking without de-anonymization and the other way around with it.
This once again highlights the importance of testing both variants as they test for different phenomena in the anonymized data.

\section{Conclusion}\label{sec:conclusion}
In this paper we presented SEBA, an easy-to-use and easy-to-extend software framework for a strong evaluation of biometric anonymizations.
We introduced our evaluation methodology which is a combination and simplification of the latest advances in related work.
We also showed how to compare anonymizations fairly by evaluating their privacy-utility trade-off using area-under-curve as a metric.
During a prototypical experiment, we find that the AUC values give a fast first impression for the anonymizations, while the privacy-utility plots allow for a more detailed comparison.

SEBA's design was purposely designed to be easy-to-use for new anonymization evaluations, which we hope to see in the future.
It is therefore an important contribution in improving the evaluation of anonymizations and therefore making them more secure.

\begin{acks}
This work was funded by the Topic Engineering Secure Systems of the Helmholtz Association (HGF) and supported by KASTEL Security Research Labs, Karlsruhe.
Funded by the German Research Foundation (DFG, Deutsche Forschungsgemeinschaft) as part of Germany’s Excellence Strategy – EXC 2050/1 – Project ID 390696704 – Cluster of Excellence “Centre for Tactile Internet with Human-in-the-Loop” (CeTI) of Technische Universität Dresden.
\end{acks}

\bibliographystyle{ACM-Reference-Format}
\bibliography{main}

\appendix

\section{Sample images}\label{sec:samples}
\begin{figure}[ht!]\centering\setlength{\tabcolsep}{2pt}
\begin{tabular}{*{3}{>{\centering\arraybackslash}m{2.55cm}}}
  \includegraphics[width=2.5cm]{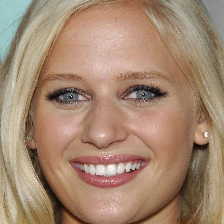} & \includegraphics[width=2.5cm]{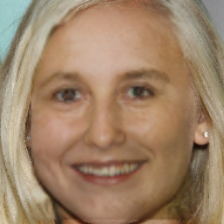} & \includegraphics[width=2.5cm]{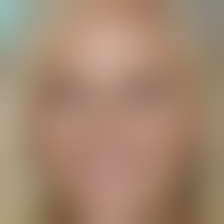} \\
  Clear & DeepPrivacy & Blurring(51)\\
  \includegraphics[width=2.5cm]{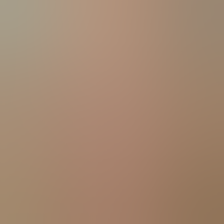} & \includegraphics[width=2.5cm]{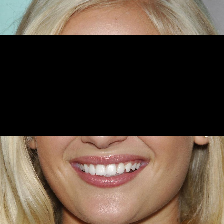} & \includegraphics[width=2.5cm]{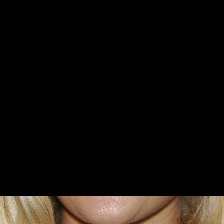} \\
  Blurring(147) & Masking(50) & Masking(110)\\
\end{tabular}
\caption{Tested anonymizations in our sample experiment}
\label{fig:sample-imgs}
\end{figure}

Sample images for the anonymizations and their parameterizations that we considered in our sample experiment can be found in Figure~\ref{fig:sample-imgs}.

\section{Implemented Methods}\label{sec:implemented}
In this section, we note all the methods already implemented in SEBA.
\subsection{Anonymizations}
\begin{itemize}
    \item Face
    \begin{itemize}
        \item Autoregressive Poisoning \cite{sandovalsegura2022autoregressive}
        \item Block Permutation
        \item CIAGAN \cite{maximov_ciagan_2020}
        \item DeepPrivacy \cite{hukkelas_deepprivacy_2019}
        \item DP Pix \cite{fan_image_2018}
        \item DP Samp \cite{fan_image_2018}
        \item DP Snow \cite{john_let_2020}
        \item Eye Masking
        \item Fawkes \cite{shan_fawkes_2020}
        \item Gaussian Blur
        \item Gaussian Noise
        \item k-RTIO \cite{rajabi_impracticality_2021}
        \item \textit{k}-Same-Eigen \cite{newton_preserving_2005}
        \item \textit{k}-Same-Pixel \cite{newton_preserving_2005}
        \item Pixelation
        \item Pixel Relocation \cite{cichowski_reversible_2011}
    \end{itemize}
    \item Gait~\cite{hanisch_understanding_2023}
    \begin{itemize}
        \item Coarsening
        \item Motion extraction
        \item Noise injection
        \item Normalization
        \item Remove parts
        \item Resampling
        \item Simple size normalization
        \item Sum of parts
        \item Trajectory feature extraction
    \end{itemize}
\end{itemize}

\subsection{De-Anonymizations}
\begin{itemize}
    \item Blind deconvolution using a normalized sparsity measure approach \cite{krishnan_blind_2011}
    \item Deep-Face Super-Resolution \cite{ma_deep_2020}
    \item Fantômas \cite{todt_fantomas_2024}
    \item Linear/Bicubic interpolation
    \item MPRNet~\cite{zamir_multi-stage_2021}
    \item Pix2Pix \cite{isola2017image}
    \item Richardson-Lucy Deconvolution \cite{richardson_bayesian-based_1972, fish_blind_1995}
    \item Stripformer \cite{tsai_stripformer_2022}
    \item Wavelet denoising~\cite{chang_adaptive_2000}
    \item Wiener filter \cite{hunt_matrix_1971}
\end{itemize}

\subsection{Recognition Methods}
\begin{itemize}
    \item Face
    \begin{itemize}
        \item ArcFace \cite{deng_arcface_2019}
        \item AWS Rekognition
        \item DeepFace \cite{serengil_lightface_2020}
        \item FR-KNN \cite{geitgey_face_2021}
        \item PCA + SVM
    \end{itemize}
    \item Gait
    \begin{itemize}
        \item PCA + SVM
        \item PCA + Random Forest
    \end{itemize}
\end{itemize}

\subsection{Utility Measures}
\begin{itemize}
    \item Face
    \begin{itemize}
        \item Attribute Similarity via DeepFace \cite{serengil_lightface_2020}
        \item Landmark Similarity via MediaPipe \cite{mediapipe}
        \item Learned Perceptual Image Patch Similarity \cite{zhang_unreasonable_2018}
        \item Structural Similarity \cite{wang_image_2004}
    \end{itemize}
    \item Gait
    \begin{itemize}
        \item Attribute classification via SVM
        \item Dynamic time warping
        \item Fréchet distance
        \item Euclidean distance
    \end{itemize}
\end{itemize}

\section{Interfaces}\label{sec:interfaces}
This section serves as a documentation which methods will need to be implemented in order to extend SEBA.

\subsection{Trait}
To add additional biometric traits, subfolders containing at least one technique for this trait will need to be created for anonymization, privacy and utility. Finally, a dataset which uses this trait will need to be added.

\subsection{Anonymization}
\begin{python}
class AbstractAnonymization:
    # Either implement this function to 
    # anonymize a single datapoint
    def anonymize(self, point):
        pass

    # Or overwrite this function to 
    # anonymize the entire data set 
    def anonymize_all(self):
        for point in self.dataset.datapoints.values():
            self.anonymize(point)
\end{python}

\subsection{De-Anonymization}
\begin{python}
class AbstractDeanonymization:
    # Train the specific De-anonymization model
    # for a give anonymization using the attacker 
    # data set
    def train(self, clear_set, anon_set):
        pass

    # Either implement this function to 
    # de-anonymize a single datapoint
    def deanonymize(self, point):
        pass

    # Or overwrite this function to 
    # anonymize the entire data set 
    def deanonymize_all(self):
        for point in self.dataset.datapoints.values():
            self.deanonymize(point)
\end{python}

\subsection{Recognition}
\begin{python}
class Classification(Inference):
    # Enrolls samples for every class
    def enroll(self, set):
        pass

    # If the classification method requires
    # training before enrolling the target classes
    # implement this function to get the background
    # data set
    def train(self, set):
        pass

    # Either implement this function to 
    # classify a single data point
    def classify_point(self, point):
        pass

    # Or overwrite this function to 
    # classify the entire data set 
    def classify_all(self, set, results):
        for name, point in set.datapoints.items():
            results.append(self.classify_point(point))
        return results
\end{python}

\subsection{Utility}
The utility can either be done using a classification which has the same interface as for the recognition or via a comparison which has the following interface.
\begin{python}
class Comparison(Inference):
    # Train the inference model on a background
    # data set
    def train(self, set):
        pass

    # Either implement this function to 
    # measure a single datapoint
    def compare_point(self, old_point, new_point):
        pass

    # Or overwrite this function to 
    # measure the entire data set 
    def compare_all(self, orig_set, new_set, results):
        for key in orig_set.datapoints.keys():
            results.append(self.compare_point 
            (orig_set.datapoints[key], 
            new_set.datapoints[key]))
\end{python}
\vspace{30em}

\end{document}